\documentclass[superscriptaddress,aps,prapplied,reprint]{revtex4-2}
\usepackage{CJK}
\usepackage{graphicx}% Include Fig. files
\usepackage{dcolumn}% Align table columns on decimal point
\usepackage{bm}% bold math
\usepackage{amssymb}
\usepackage{amsmath}
\usepackage{wasysym}
\usepackage{physics}
\usepackage[version=4]{mhchem}
\usepackage{float}
\usepackage{color}
\usepackage{enumitem}
\usepackage{subcaption,booktabs}
\usepackage{adjustbox}
\usepackage{ulem}
\makeatletter
\renewcommand\sout[1]{\@bsphack\@esphack}%
\makeatother

\newcommand{\rev}[1]{\textcolor{black}{#1}}

\newcommand{\zstroke}{%
  \text{\ooalign{\hidewidth -\kern-.3em-\hidewidth\cr$z$\cr}}%
}

\begin{document}

\title{Electroconvective instability in water electrolysis: an evaluation of electroconvective patterns and their onset features}

\author{Nakul Pande}
\affiliation{Physics of Fluids, University of Twente, Enschede, NL}
\affiliation{PhotoCatalytic Synthesis, University of Twente, Enschede, NL}

\author{Jeffery A. Wood}
\affiliation{Soft matter, Fluidics and Interfaces, University of Twente, Enschede, NL}

\author{Guido Mul}
\affiliation{PhotoCatalytic Synthesis, University of Twente, Enschede, NL}
\author{Detlef Lohse}
\affiliation{Physics of Fluids, University of Twente, Enschede, NL}

\author{Bastian T. Mei}
\affiliation{PhotoCatalytic Synthesis, University of Twente, Enschede, NL}
\author{Dominik Krug}
\affiliation{Physics of Fluids, University of Twente, Enschede, NL}
\noaffiliation
\date{\today}

\begin{abstract}
In electrochemical systems, an understanding of the underlying transport processes is required to aid in their better design. This includes knowledge of possible near-electrode convective mixing that can enhance measured currents. Here, for a binary acidic electrolyte in contact with a platinum electrode, we provide evidence of electroconvective instability during electrocatalytic proton reduction. The current-voltage characteristics indicate that electroconvection, visualized with a fluorescent dye, drives current densities larger than the diffusion transport limit. The onset and transition times of the instability do not follow the expected inverse-square dependence on the current density, but, above a bulk-reaction-limited current density are delayed by the water dissociation reaction, \rev{i.e. the formation of $\text{H}^+$ and $\text{OH}^-$ ions}. The dominant size of the electroconvective patterns is also measured and found to vary \sout{as} \rev{with} the diffusion length scale, confirming previous predictions on the size of electroconvective vortices. 
\end{abstract}

\maketitle

\section{Introduction}
Electrolysis is projected to be a core-technology for a sustainable society \cite{Yan2020}, with applications in energy storage (lithium-ion batteries, water electrolysis), climate change mitigation ($\text{CO}_2$ reduction), and production of useful chemicals (selective hydrogenation, $\text{N}_2$ reduction to $\text{NH}_3$). Economical considerations typically require electrolyzers for these processes to operate at large current densities \cite{LeRoy1983,Burdyny2019}, at which mostly transport processes are rate-limiting. 
This triggers special interest into convective phenomena as a possible driver of `overlimiting currents', i.e. beyond the diffusive ion-transport limit. In the absence of external mixing, such convection may be induced by buoyancy \cite{Wagner1949,Ngamchuea2015}, but can interestingly also originate from electrohydrodynamic forces. The latter, known as electroconvection, is generally associated with the instability at ion-selective interfaces (membranes or electrodes) of the formed space-charge layer \cite{Rubinstein2000,Druzgalski2016,Mani2020}. Electroconvection has been studied extensively on ion-exchange membranes (IEM) in the context of water desalination \cite{Rubinstein2008,Yossifon2008,Kang2020,Nikonenko2014}, on electrodes for metal electrodeposition \cite{Fleury1992,Fleury1993,Huth1995,Bai2016,Zhang2020} \rev{, and recently on inert electrodes subjected to AC-voltage \cite{Kim2019}}. Particularly \rev{for electrodeposition}, it has been shown \cite{Fleury1992,Fleury1993,Huth1995} that electroconvection leads to a change in the morphology of the metal deposit and dendrite formation, which may result in short-circuiting in lithium-ion batteries and has implications on their design \cite{Bai2016}. 
However, the electro-catalytic processes of water electrolysis, $\text{CO}_2$ and $\text{N}_2$ reduction additionally involve non-linear bulk reactions. Their presence is known to  have a strong influence on the pH distribution \cite{Andersen2014a,Pande2020} and fluid properties \cite{Ngamchuea2015,Obata2020,DeWit2020}, but how this affects the electroconvective phenomenon in these important systems remains unclear. 

Electrohydrodynamic patterns have been reported previously in some electrolytic systems, e.g. in the electrochemiluminescence of rubrene in the non-aqueous electrolyte 1,2-dimethoxyethane \cite{Kostlin1980,Orlik1998,Orlik1998a}. In this case, the rubrene cations and anions which are formed at their respective electrodes, recombine in the electrolyte and emit light, thereby making the patterns visible. These were reported first by \citeauthor{Kostlin1980} \cite{Kostlin1980} and then by \citeauthor{Orlik1998} \cite{Orlik1998} who later presented a theoretical model based on ion-transport to explain their results \cite{Orlik1998a}. In aqueous electrolytic systems, electrohydrodynamic patterns have been visualized previously in water electrolysis using either charged carbon nanotubes \cite{Sato2007} or colloidal spheres \cite{Han2006} as tracers. By further observing patterns of oxidation on ITO electrodes in the absence of the charged colloidal spheres, \citeauthor{Han2006} \cite{Han2006} confirmed that the patterns were in fact formed by electroconvection. However, despite these early measurements, a systematic study of electroconvective patterns and their dependence on the electrical forcing is still required.

In this paper we present measurements of electroconvective patterns in water electrolysis. Details on the experimental setup employed are provided in Section \ref{sec:Exp}. In Section \ref{sec:LSV} and Section \ref{sec:CP}, we show the results of linear sweep voltammetric and chronopotentiometric experiments, respectively. An effective reaction-diffusion model derived in Appendix \ref{sec:numerics} is used to explain the experimental transition times. Finally, we summarize our findings in Section \ref{sec:conclusion}.

\section{Experimental Setup}\label{sec:Exp}
\begin{figure*}
\includegraphics{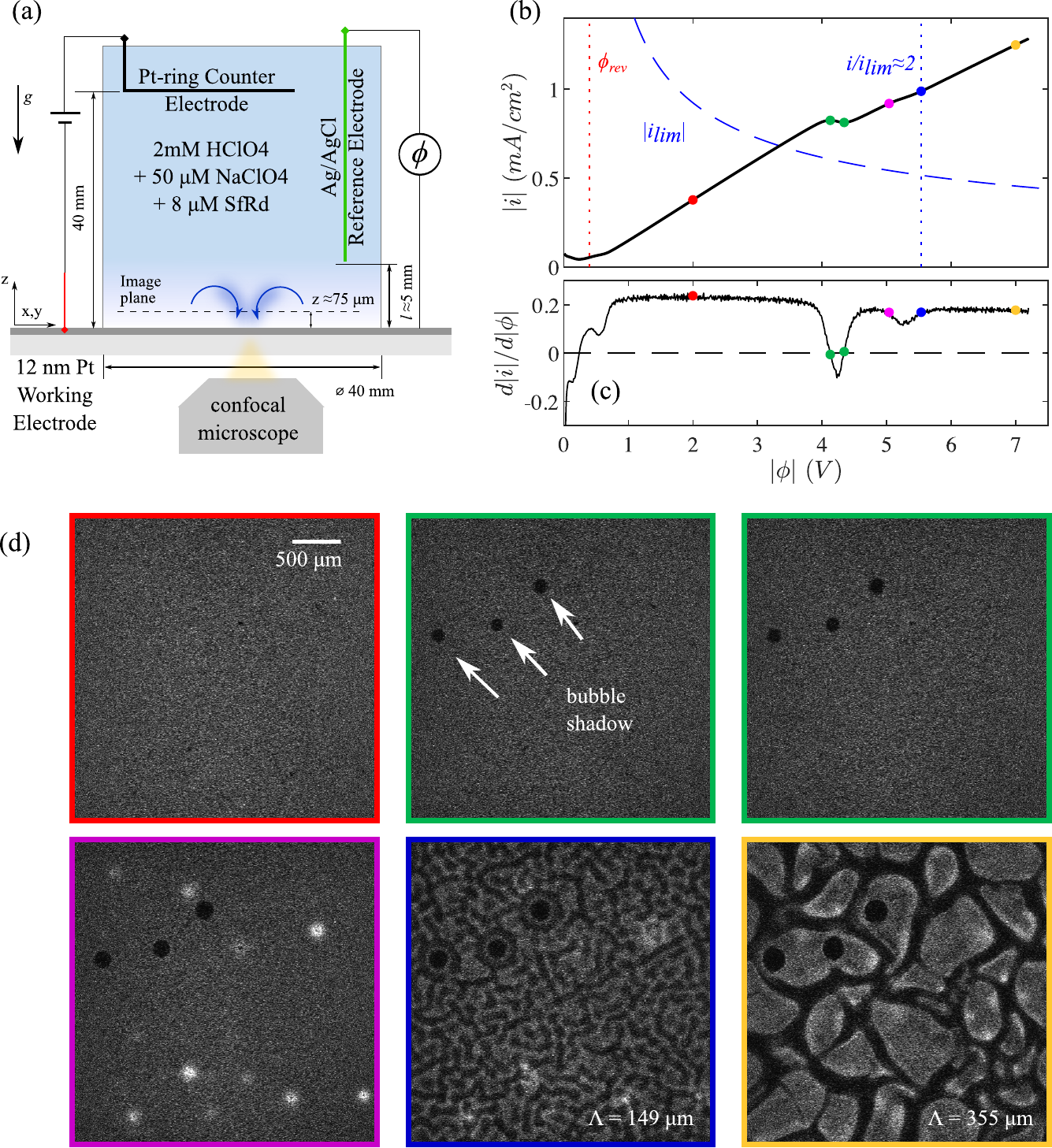}
\caption{\label{fig:CV}(a) Schematic of the experimental setup. (b) A typical linear sweep voltammogram representative of the system response measured at a rate $\abs{d\phi/dt}=0.1~\text{V/s}$. The limiting current based on the Cottrell equation $i_{lim}(t)$ has been plotted as a dashed blue line. The green ($d\abs{i}/d\abs{\phi} = 0$), magenta and blue markers (both $d^2\abs{i}/d\abs{\phi}^2 = 0$) have been placed at locations of the respective transitions in $\abs{i}$ as shown in (c). The fluorescence images in (d) are measured simultaneously with the linear sweep, and the color code corresponds to the markers in (b). The dominant wavelength, $\Lambda$, of the patterns in (d) is calculated as described in Section \ref{sec:VortSize}}
\end{figure*}

A schematic of the employed setup with relevant dimensions is shown in Fig. \ref{fig:CV}(a). We use a cylindrical electrochemical cell made of Teflon with a transparent platinum working electrode and a Pt-ring mesh counter electrode (distance between working and counter electrode $\approx4$ cm) for the measurements. The transparency of the platinum electrode was obtained by sputtering a 12 nm platinum layer (with a 3 nm chromium under-layer for better adhesion) on a 170 $\mu$m glass slide. Electrical contact to the thin film platinum electrode was made using a platinum contact pressed onto the electrode. The electrical measurements were made with a VersaStat potentiostat using a Ag/AgCl (BasiR) reference electrode. Overall, the experimental design allows for a quasi 1-D \rev{\footnote{Here, the counter electrode, (also diameter, $\diameter = 4$ cm) is a ring electrode (radial symmetry; and a thickness of $\approx 1$ mm), and is placed $4$ cm away from the working electrode. This distance is much larger than the diffusion length scale of the system for $\sqrt{D_S t} \sim 3~\text{mm}$. We therefore expect no influence of the counter electrode on our results.}} and unrestricted progression of the depletion front. This is in contrast to previous measurements which were performed in thin-layer electrochemical cells \cite{Kostlin1980,Orlik1998,Orlik1998a,Han2006,Sato2007} (distance between electrodes $\sim 100~\mu\text{m}$) where the interaction overlapping concentration boundary layers and bulk-recombination reactions could make the analysis of the system complicated. 

The electrochemical cell was mounted on an inverted laser scanning confocal fluorescent microscope (Nikon confocal microscope A1 system, Nikon Corporation, Tokyo, Japan) with a 4x dry objective (CFI Plan Fluor 4x/0.13) which was used to measure a 3.17 mm $\times$ 3.17 mm region (512 $\times$ 512 $\text{pixel}^2$) chosen close to the center of the electrode. A 561 nm excitation laser was used to excite the chosen fluorescent dye (Sulforhodamine 101), while the emission was collected in a $545-645$ nm wavelength window with a pinhole size of 28.1 $\mu$m. The fluorescence measurements were made at the z-location of maximum fluorescence intensity which was $z\approx 75~\mu$m above the electrode surface. The electrode surface was in turn found by the maxima of the reflected light intensity \cite{Pande2020}. Sulforhodamine 101 was chosen for measurements since it is a pH and temperature insensitive dye \cite{Coppeta1998}. A relatively small concentration of 8 $\mu$M ensured that self-quenching of its fluorescence signal, which is observed at dye concentrations $\approx 100~\text{mM}$ \cite{Scida2019}, was avoided. We further assume electrochemical stability of the dye during the (reductive) measurement. For all the measurements, the electrolyte was composed of 2 mM $\text{HClO}_4$, 8 $\mu \text{M}$ of Sulforhodamine 101, and 50 $\mu$M of supporting salt $\text{NaClO}_4$. All chemicals were obtained from Sigma-Aldrich. Note that the supporting or indifferent salt affects the conductivity of the electrolyte (electrical migration effects in solution), and was purposely avoided to create conditions suitable for the onset of  electroconvection. However, the application of the largest current density considered in this work ($\approx 10~\text{mA/cm}^2$) required the addition of at least 50 $\mu$M of $\text{NaClO}_4$. Nevertheless, the concentration of the supporting $\text{Na}^+$ ions, $c_{\text{Na}^+}$ is much smaller than that of the reacting $\text{H}^+$ ion, $c_{\text{H}^+}$, i.e.  $c_{\text{Na}^+}\ll c_{\text{H}^+}$ such that a binary electrolyte approximation can be used.

\section{Results and Discussion}
\subsection{Linear sweep voltammetry}\label{sec:LSV}
We first measure the current-voltage relationship, since the onset of electroconvection at IEMs is typically associated with a characteristic shape of the current-potential curve \cite{Rubinstein1979a,Rubinstein2008,Kang2020}. To examine this behavior for electro-catalytic proton reduction, we sweep the potential (at a rate of $v = 0.1~\text{V/s}$, starting at $0~\text{V}$), as is common practice in electrochemistry \cite{BardFaulkner2002}. The derived voltage-dependent current density, $\abs{i(\phi)}$ is plotted in Fig. \ref{fig:CV}(b). A small decrease in $\abs{i}$, associated with a transient capacitive current \cite{BardFaulkner2002} (due to sudden potential change from the open circuit potential, $\sim400~\text{mV}$, to start potential of the sweep) is followed by a continuous increase up to $\abs{\phi}\lessapprox 4$V. After an additional decrease in $\abs{i}$, the current resumes its increasing trend with increasing $\abs{\phi}$. This behavior is similar to that encountered on IEMs. In that case, the
transient levelling-off is associated with reaching the diffusion limit and the subsequent recovery of an increasing slope in the $\abs{i(\phi)}$ curve is linked to the driving of overlimiting currents due to electroconvective transport \cite{Rubinstein2008,Kang2020}. This analogy suggests that the transition indicated by the green markers in Fig. \ref{fig:CV}(b) is related to the diffusion limit of proton reduction at the electrode surface and electroconvection setting in for $\phi>4\text{V}$. The latter is confirmed by the results of the fluorescence imaging in Fig. \ref{fig:CV}(d). These images were measured simultaneously with the voltammogram and the corresponding times are indicated with markers of same colours in Fig. \ref{fig:CV}(b) and \ref{fig:CV}(c). Specifically, green markers are placed at $d\abs{i}/d\abs{\phi} = 0$ and magenta and blue markers indicate points at which $d^2\abs{i}/d\abs{\phi}^2 = 0$ (see Fig. \ref{fig:CV}(c)).
Since the fluorescent dye molecules are \rev{negatively (likely single) charged \rev{\cite{Yaguchi2019} }}, the lateral electric-field gradients \rev{(parallel to the electrode)} associated with electroconvection, i.e. due to the instability of the polarized layers near the electrode \cite{Druzgalski2016}, \rev{initially} lead to inhomogeneities in the image plane. \rev{In Fig. \ref{fig:CV}(d),} \rev{i}nitially (red marker), the dye distrbution is homogeneous and besides the appearance of bubble shadows the image remains unchanged even as the potential is increased. At $\phi\approx 5$V, the first inhomogeneity in the dye distribution appears (magenta marker), and eventually a very distinct pattern emerges over the entire imaged area (blue marker). At later times and even higher potentials (yellow), the typical pattern size is significantly larger, indicating an increase in size of the electroconvective vortices. \rev{The appearance of the patterns and its relation to fluid motion is likely as follows. 
Initially, the current at the electrode drives a depletion of ions near its surface. After significant depletion, at the onset of the instability, electric field driven forces appear in directions parallel to the electrode creating 1) an in-plane inhomogeneity in the fluorescent image, and 2) a non-uniform electro-osmotic slip due to its action on the space-charge layer. The resulting vortices bring in fresh fluid (rich in both positive and negative ions) from the bulk. This region of downward flow on to the electrode likely induce the bright spots in the image (as  results of \citeauthor{Davidson2016} \cite{Davidson2016} indicate). At the same time, the vortices also deplete the imaged plane of ions, corresponding to the dark spots at some locations in the image. Therefore the patterns are related to the influx or efflux arm of the vortex. Ultimately, since the ions of a molecule could be thought of as migrating in cation-anion pairs, the concentration (or intensity) of the fluorescence dye should be proportional to the concentration of the background binary acid.}

Determining the ratio of the measured and the limiting currents reveals to what extent the emergence of convective structures is associated with overlimiting currents. \rev{An estimate} of \sout{The} \rev{the} diffusion-limited current density \sout{$i_{lim}$} can be \sout{estimated using} \rev{obtained from} the Cottrell equation, \rev{which holds in case the applied  potential step is sufficiently large to reduce the concentration of the electroactive species (here $\text{H}^+$ ions) to zero at the electrode surface (i.e. in diffusion limited conditions)} \cite{BardFaulkner2002}. \rev{Specifically, the transient diffusion equation is solved in a semi-infinite domain with a zero concentration boundary condition at the electrode, i.e. $c_{\text{H}^+}(0,t>0) = 0$ to yield the time-dependent current density $i$. In the case of a binary electrolyte this results in}
\begin{equation}
    i_{lim}(t) = \rev{-FD_s\frac{c_0}{(1-\tau_{\text{H}^+})\sqrt{\pi D_st}}}, \label{cottrell}
\end{equation}
where $F = 96485~\text{C/mol}$ is Faraday's constant, \sout{$A$ is the electrode area,} $D_s = \frac{2D_{\text{H}^+}D_{\text{ClO}_4^-}}{D_{\text{H}^+}+D_{\text{ClO}_4^-}}$ is the effective salt diffusivity, $\tau_{\text{H}^+} = \frac{D_{\text{H}^+}}{D_{\text{H}^+}+D_{\text{ClO}_4^-}}$ is the ion transport number for the $\text{H}^+$ ion and $c_0$ the bulk concentration of the $\text{H}^+$ ion. $D_{\text{H}^+}$ and $D_{\text{ClO}_4^-}$ are the diffusivities of the $\text{H}^+$ and $\text{ClO}_4^-$ ions, respectively. \sout{Specifically, $i_{lim}(t)$ is a time-dependent function obtained by solving the transient diffusion equation in a semi-infinite domain with a zero concentration boundary condition, for a binary electrolyte.} 
\rev{We note that a zero concentration of $\text{H}^+$ ions at the electrode can only occur at potentials larger than the reversible potential for hydrogen evolution $\phi_{rev}$ (i.e. $0~V$ vs R.H.E. reference; calculated to be $\phi_{rev}= -0.3972~V$ for Ag/AgCl reference in Appendix \ref{sec:PotDeb}). Therefore equation \eqref{cottrell} is applied from the time when $\abs{\phi} > \abs{\phi_{rev}}$, i.e. starting at $t = t_{\abs{\phi} = \abs{\phi_{rev}}}$ ($t = 0$ in equation \eqref{cottrell} is in fact $t = t_{\abs{\phi} = \abs{\phi_{rev}}}$).} It is shown in Fig. \ref{fig:CV}(b) that pattern formation occurs at $i/i_{lim}\approx 2$, meaning that the pattern formation observed in Fig. \ref{fig:CV}(d) does indeed arise at overlimiting currents. \rev{A similar conclusion can be drawn from a further comparison of the current-voltage measurements with analytic expressions that take the time-dependence of the potential into account. This is presented in Appendix \ref{sec:LSV_norm} where, additionally, the existence of overlimiting currents in water electrolysis at higher supporting electrolyte concentration is discussed.}
% \rev{Similar conclusions can be drawn from a further comparison of the current-voltage measurements with analytic expressions that take the time-dependence of the potential into account as presented in Appendix \ref{sec:LSV_norm}.
% we also compare in Fi. .. with different supporting electrolytes  overlimiting currents are a feature of water electrolysis. In a binary electrolyte where electroconvection is clearly the driver}

% This could be completed, for example, by noting that the system very quickly becomes diffusion-limited. Another point of comparison that the authors should consider is an analytic expression for linear sweep voltammograms. Bard and Faulkner (among others) provide a solution demonstrating the expected nonmonotonic behavior, with a well-defined peak and subsequent diffusion-limited decay at higher potentials, and such a solution might be adapted to this present work. It would be insightful to see how this analytic result compares to the authors’ measurements, and where exactly the measured current diverges from the analytic expression.
%\rev{However, it is fair to say that a zero concentration of $\text{H}^+$ ions at the electrode occurs only for potentials greater than the reversible potential for hydrogen evolution $\phi_{rev}$ ($0~V$ vs R.H.E. reference; calculated for Ag/AgCl reference in Appendix \ref{sec:PotDeb}). Therefore equation \eqref{cottrell} can only be applied from the time when $\phi > \phi_{rev} = -0.3972~V$, i.e.starting at $t = t_{\phi = \phi_{rev}}$ ($t = 0$ in equation \eqref{cottrell} is in fact $t = t_{\phi = \phi_{rev}}$)} 

Note that the analysis presented \rev{in this section, does not include the role of the water bulk reaction.} \sout{here} \rev{The analysis presented in this paper, moreover,} does not include the Second Wien effect \cite{Onsager1934}, i.e. the increase in the rate of production of H$^+$ ions due to an enhancement of the dissociation constant $k_b$ for water at high electric-fields \cite{Tanaka2010} (for chronopotentiometric measurements an estimate of this effect is provided in Appendix \ref{sec:PotDeb}). However, the Second Wien effect will likely also play a role here and may be one of the reasons why electroconvection only sets in for $i/i_{lim}> 1$. Thus, there is likely an additional mechanism of enhanced dissociation rate which could provide additional reacting ions to support the increase in $\abs{i}$ with $\abs{\phi}$.
\begin{figure}
\includegraphics{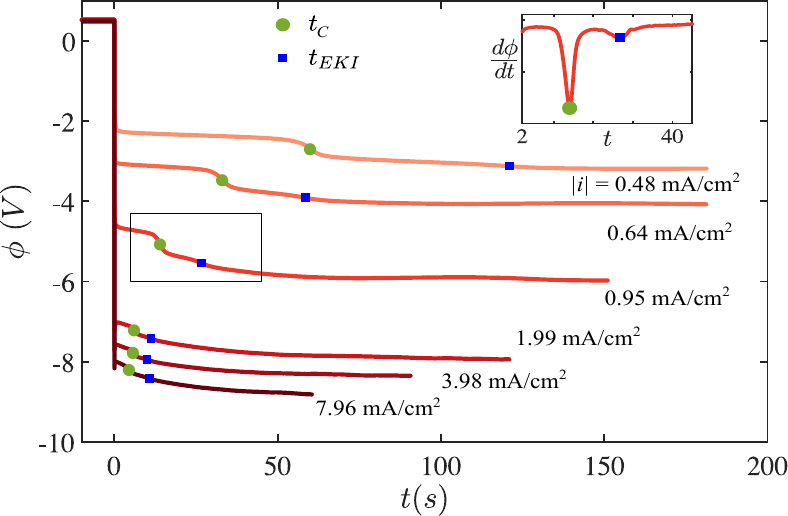}
\caption{\label{fig:constCurrent_1} Measured electrode potential (vs Ag/AgCl) at various constant current densities $\abs{i}$. The markers indicate locations where $d^2\phi/dt^2=0$ and the inset shows $d\phi/dt$ for $\abs{i}= 0.95$ mA/cm$^2$ (corresponding to the framed box in the interval $2~\text{s}\leq t \leq 45~\text{s}$).}
\end{figure}

\subsection{Chronopotentiometric measurements}\label{sec:CP}
While voltammetric experiments are consistent with electroconvection, detailed analysis is complicated given that both $i$ and $\phi$ vary in time. To correlate the onset of electroconvection to changes in the measured potential \cite{DeValenca2015,DeValenca2017}, we therefore performed experiments at constant current densities. For such configurations, the transition to electroconvection has further been linked to Sand's time, i.e. the time at which the electrolyte concentration vanishes at the boundary \cite{BardFaulkner2002}. Thus, using constant-current experiments the transition times in $\phi$, the pattern onset time $t_{ons}$, and Sand's time can be correlated.

\subsubsection{Characteristic times}

\begin{figure*}
\includegraphics{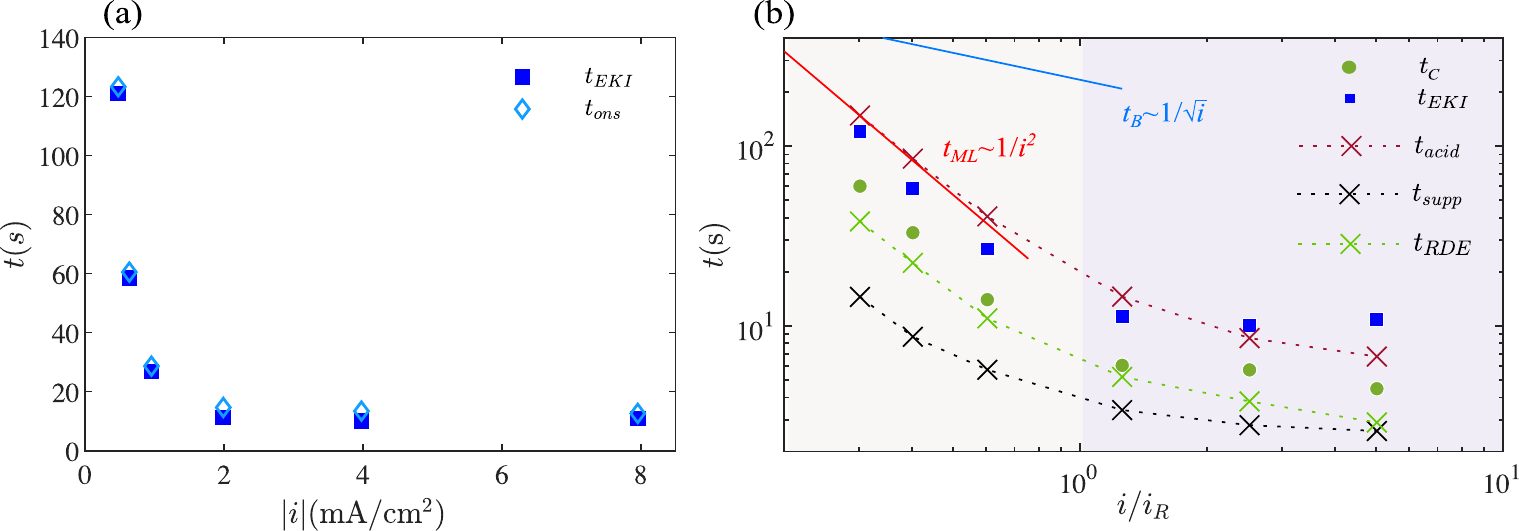}
\caption{\label{fig:constCurrent_2} \rev{(a)} \sout{The inset} shows the onset time of the patterns from the images $t_{ons}$, compared to the transition time $t_{EKI}$ measured from Fig. \ref{fig:constCurrent_1}. \rev{(b)} Log-Log plot of the transition times $t_{EKI}$ and $t_{C}$ vs. $i$ normalized by the bulk-reaction-limited current density $i_{R}$. \rev{The error bars are smaller than the symbols\footnote{The uncertainty in the measurement of the experimental times $t_C$ and $t_{EKI}$ is related to the window size of the moving average filter (50 points) that is applied to obtain $d\phi/dt$ shown in the inset of \ref{fig:constCurrent_1} (a). Therefore, the error bars are $\pm 0.25$ s. For $t_{ons}$ the error bars are related to the time resolution of the optical measurement, i.e. $\pm 0.033$ s}.} $t_{ML}$ is Sand's time as defined in \cite{Morris1963}.
$t_{acid}$, $t_{supp}$ and $t_{RDE}$ are transition times ($c\rightarrow0$) discussed in the main text and obtained from numerical simulation of the reaction-diffusion system of equations defined in Appendix \ref{sec:numerics}. Each of these are depicted with a cross marker and provided with dotted lines to guide the eye. $t_{B}$ is the onset time for buoyancy convection using equation \eqref{eq:Ra1}. \rev{Note the convention: solid symbols indicate experimental times, the oblique cross symbols the numerical times, and the solid lines the analytical estimates.}}
\end{figure*}

The potentials recorded during the constant-current experiments are presented in Fig. \ref{fig:constCurrent_1}. All considered values of the current density $\abs{i}$ are at least a 100 times greater in magnitude than the steady-state diffusion limited current density, $i_{lim} = -FD_Sc_0/L = -2.89~\mu\text{A/cm}^2$, where $L = 20$ mm is half the distance between the working and counter electrode (see Fig. \ref{fig:CV}(a)). The fluorescence images (see Appendix, Fig. \ref{fig:ImgAllCur}) confirm that pattern formation is observed in all experiments. From the potential curves, we identify two distinct times, which correspond to local minima in $d\phi/dt$ as shown in the inset of Fig. \ref{fig:constCurrent_1}. From Fig. \ref{fig:constCurrent_2}(a), the later of the two times (blue square) is found to be in very good agreement with the onset time of pattern formation, $t_{ons}$, (obtained by visual inspection of the dye images), at all $\abs{i}$. This connects the second (less pronounced) minimum in $d\phi/dt$ to the onset of electrokinetic instability which is therefore referred to as $t_{EKI}$ in the following. The first minimum in $d\phi/dt$, named $t_C$, is not observed on IEMs \cite{DeValenca2017} and is likely related to a change in the reaction at the electrode due to transport limitations, as will be shown later.

Before doing so, it is useful to consider the dependence of $t_{EKI}$ and $t_C$ on $\abs{i}$. First, we compare our experimental observations in Fig. \ref{fig:constCurrent_2}(b) to an estimate for Sand's time given by \citeauthor{Morris1963} \cite{Morris1963}, $t_{ML} = \frac{\pi D_s}{4}\left(\frac{c_0F}{(1-\tau_{\text{H}^+})i}\right)^2$ (red line), which applies in the absence of a bulk reaction and for a binary electrolyte assuming electroneutrality. For small current densities, $t_{EKI}$ is found to be in good agreement with $t_{ML}$. Moreover, both $t_C$ and $t_{EKI}$ follow the classical $1/i^2$ relationship implied by the expression for $t_{ML}$. With increasing $\abs{i}$ however, the decrease of $t_C$ and $t_{EKI}$ is significantly slower. In fact, we find that the deviation occurs around a ``bulk reaction-limited'' current density $i_R$, which reflects the maximum current (or proton flux) that can be sustained at the electrode by the bulk-reaction term when the concentration at the electrode reaches diffusion limitation. In particular, $i_R= -FD_{\text{H}^+}c_0/\delta_R = -Fk_bc_W\delta_R$, where, $\delta_R = \sqrt{\frac{D_{\text{H}^+}c_0}{k_bc_W}}$ is the size of the reaction-diffusion boundary layer for dissociation of water, based on a balance between diffusive flux $\sim D_{\text{H}^+}\frac{c_0}{\delta_R^2}$ and production due to water dissociation $\sim k_bc_W$, where $k_b$ and $c_W$ are the dissociation constant and water concentration respectively (values of constants can be found in Appendix \ref{sec:numerics}). 

To explore the reason for the change in scaling behavior for $t_C$ and $t_{EKI}$ at $i/i_R\approx 1$ (see Fig. \ref{fig:constCurrent_2})(b), we consider the ion transport equations for the different components of the acidic electrolyte. As shown in Appendix \ref{sec:numerics}, assuming electroneutrality, the corresponding system of equations is reduced to a reaction-diffusion one in the limit of two extreme cases: 1) in the absence of supporting electrolyte, called $acid$ henceforth, and 2) with excess supporting electrolyte ($supp$). While in $acis$ the transport of $\text{H}^+$ ions is coupled to $\text{ClO}_4^-$ ions, in $supp$, the $\text{H}^+$ ions move independently of the other ions in solution. For both cases, we integrate the transport equations numerically and obtain the respective Sand's times, $t_{acid}$ and $t_{supp}$, which are presented in Fig. \ref{fig:constCurrent_2}(b). $t_{acid}$ and $t_{supp}$ provide an envelope for both the experimental transition times, i.e.  $t_{supp}<t_C,t_{EKI}<t_{acid}$. Moreover, $t_{acid}$, which is an estimate of Sand's time in the case of negligible supporting electrolyte, approximates $t_{EKI}$ reasonably well. Importantly, $t_{acid}$ approaches $t_{ML}$ (see previous paragraph for definition) at $i/i_R\ll 1$, and differs significantly from $t_{ML}$ for $i/i_R \gtrapprox 1$, revealing the increasing importance of the bulk-reaction at higher current densities, which causes the deviation from the classical scaling. 

The preceding discussion has shown that the time of the second (electroconvective) transition in $\phi$ is approximated reasonably well by Sand's time $t_{acid}$. The small but distinct potential change at $t_{EKI}$ in Fig. \ref{fig:constCurrent_1} is therefore related to the vanishing conductivity at the boundary (see equation \eqref{eq:phiGrad}). Here, we provide a possible explanation for the additional transition at $t_C$ in terms of a change in the reaction at the electrode at the moment when diffusion limitation i.e. $c_{\text{H}^+}\approx0$ is encountered at the electrode surface. 

To do so, we consider the kinetic boundary condition at the reaction plane which can be expressed as the generalized Frumkin-Butler-Volmer equation \cite{Bonnefont2001, VanSoestbergen2010, VanSoestbergen2012, Yan2017} for the Faradaic current ($i_F$) at the boundary. It reads

\begin{subequations}
\begin{align}
    i_F &= k_{-}c_{\text{H}^+}\Gamma(1-\theta) e^{-\alpha\Delta\phi_S/\phi_T} - k_{+}\Gamma \theta e^{\alpha\Delta\phi_S/\phi_T}\label{eq:BV}\\
    \text{where,}\nonumber\\
    i_F &= D_{\text{H}^+}\left(\pdv{c_{\text{H}^+}}{z} + \frac{c_{\text{H}^+}}{\phi_T}\pdv{\phi}{z}\right)\eval_{z = 0}. \label{eq:iF}
\end{align}
\end{subequations}
Here, $k_{-}$ is the rate constant for the reduction reaction (formation of $\text{PtH}_{ads}$), $k_{+}$ for the corresponding oxidation reaction, $\alpha = 0.5$, $\Gamma$ the total number of available sites for reaction, $\theta$ the fractional coverage, $\Delta \phi_S$ the potential drop across the Stern layer driving the reaction and $\phi_T = RT/F$ the thermal voltage ($R$ is the ideal gas constant, $T$ = 298 K).

We rewrite equation \eqref{eq:BV} as
\begin{subequations}
\begin{align}
    i_F &= i_{kin}\left(1 - \frac{k_{+}\theta}{k_{-}(1-\theta)c_{\text{H}^+}} e^{2\alpha\Delta\phi_S/\phi_T}\right) \label{eq:BV1}\\
    \text{with, }i_{kin} &= k_{-}c_{\text{H}^+}\Gamma(1-\theta) e^{-\alpha\Delta\phi_S/\phi_T},
\end{align}
\end{subequations}
where $i_{kin}$ is the kinetic current. For fast reversible reactions, such as proton reduction on platinum, $i_F/i_{kin}\ll 1$ \cite{Neyerlin2007} and quasi-equilibrium  is maintained (right hand side of equation \ref{eq:BV1} = 0). Thus, equation \eqref{eq:BV} reduces to the Nernst equation and the potential drop across the Stern layer, $\Delta\phi_S$, can be estimated using
\begin{equation}
    \frac{\Delta \phi_S}{\phi_T} = \frac{1}{2\alpha}\ln{\frac{k_{-}(1-\theta)c_{\text{H}^+}}{k_+\theta}}.
\end{equation}
Finally, it can be seen that $c_{\text{H}^+}\rightarrow 0$ is equivalent to $\theta \rightarrow 1$ at the reaction boundary, i.e. a change from the underpotential to overpotential hydrogen deposition at the electrode \cite{Conway2002}.

To estimate the time of this transition from the transport equations, we heuristically assume that the potential gradient vanishes at the reaction plane, $\pdv{\phi}{z}\eval_{z = 0} = 0$, and solve the electroneutrality equations with a modified flux boundary condition at the electrode, $D_{\text{H}^+}\pdv{c_{\text{H}^+}}{z} = \frac{i}{F}$ (instead of equation \eqref{eq:HBoundary}). This is similar to the ``zero-field approximation'' that has been used to model electrode reactions in the absence of a supporting electrolyte for fast electron transfer reactions \cite{Streeter2008,Dickinson2010}, but with the added simplification of electroneutrality. The ``zero-field approximation'' inherently assumes that the Stern layer adsorbs ions that screen the surface charge \cite{Yan2017}. Since electro-catalytic proton reduction on platinum is a fast reaction and proceeds first by an (underpotential) adsorption of protons at the boundary, this approximation appears justified in the present system. 

The transition time within this approximation, $t_{RDE}$, is the moment when the flux of protons at the electrode becomes diffusion limited i.e. when $c(z=0)=0$. Model results for $t_{RDE}$ are compared to $t_C$ in Fig. \ref{fig:constCurrent_2}(b). The values are close and the small difference between the two is indeed expected since $t_C$ corresponds to the inflection point in $\phi$ (inset Fig. \ref{fig:constCurrent_1}) whereas $t_{RDE}$ actually predicts the time at the base of the spike in $d\phi/dt$.

\begin{figure}
\centering
\includegraphics{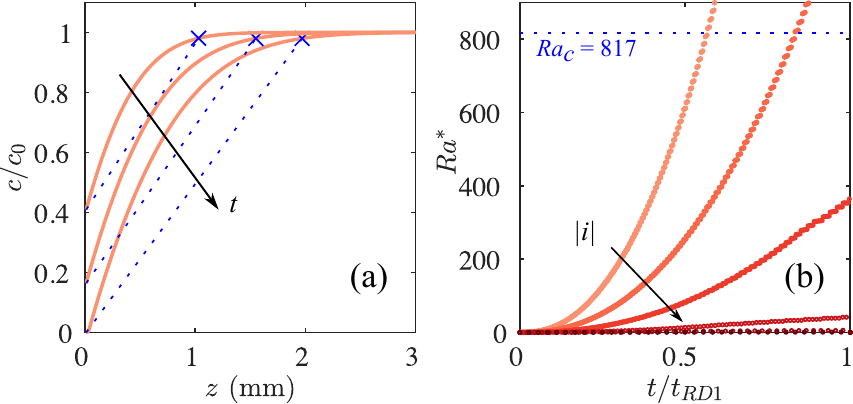}
\caption{\label{fig:buoyancy} (a) Solid lines: concentration profiles of $\text{HClO}_4$ at times $t = 51~\text{s},~103~\text{s}~\text{and}~153~\text{s}$ obtained from the numerical model '$acid$' for the smallest current density $\abs{i} = 0.48~\text{mA/cm}^2$. The dotted line shows the effective diffusion layer which is used to calculate $Ra^*$ (b) The evolution of $Ra^*$ for all current densities. $Ra_c$ is an estimate of the critical Rayleigh number for the present configuration.}
\end{figure}

\begin{table*}
%  \begin{adjustbox}{max width=\columnwidth}
\caption{\rev{Summary of the characteristic times}}
 \begin{subtable}{\textwidth}
 \caption{\rev{Experimental measured times}}
    \rev{\begin{tabular}{p{0.1\textwidth} | p{0.2\textwidth} | p{0.6\textwidth}}
    \hline
     $t_{C}$ & critical time & first peak in $d\phi/dt = 0$ in Fig. \ref{fig:constCurrent_1}, or the first time of transition in the measured electrode potential \\
     \hline
     $t_{EKI}$ & electro-kinetic instability time  & second peak in $d\phi/dt = 0$ in Fig. \ref{fig:constCurrent_1}, or the second time of transition in the measured electrode potential \\
     \hline
     $t_{ons}$ & pattern onset time & time at which a peak is detected in the radially averaged fourier spectrum of the fluorescence image, i.e. $\psi > 0.15$ in Fig. \ref{fig:waveL}(a)\\
     \hline
    \end{tabular}}
    \end{subtable}
    
    \vspace{5mm}
    
    \begin{subtable}{\textwidth}
    \caption{\rev{Numerical times}}
    \rev{\begin{tabular}{p{0.1\textwidth} | p{0.2\textwidth} | p{0.6\textwidth}}
    \hline
     $t_{acid}$ & binary acid time & time at which $c_H \rightarrow 0$ for a binary acidic electrolyte with a bulk reaction term. This serves as an estimate for the onset times of electroconvection $t_{EKI}$ (or $t_{ons}$)\\
     \hline
     $t_{supp}$  & excess supporting electrolyte time & time at which $c_H \rightarrow 0$ for an electrolyte with excess supporting salt (no potential drop in bulk solution) \\
     \hline
     $t_{RDE}$ & Reaction-Diffusion (electrode) time & time at which $c_H \rightarrow 0$, based on a model that assumes a vanishing potential gradient at the reaction plane. This serves as an estimate for the measured critical time $t_C$ and likely represents a transition from an underpotential to overpotential deposition of $\text{H}^+$.\\
     \hline
    \end{tabular}}
    \end{subtable}
    
    \vspace{5mm}
    
    \begin{subtable}{\textwidth}
    \caption{\rev{Analytical times}}
    \rev{\begin{tabular}{p{0.1\textwidth} | p{0.2\textwidth} | p{0.6\textwidth}}
    \hline
     $t_{ML}$ & Morris-Lingane time & Analytical estimate of Sand's time for purely diffusive case (no bulk reaction) as obtained in \cite{Morris1963}\\
     \hline
     $t_{B}$ & Buoyancy time & estimate for onset time for buoyancy driven convection: the time at which the Rayleigh number (buoyancy driving) exceeds critical value of 817. See equation \eqref{eq:Ra1}\\
     \hline
    \end{tabular}}
    \end{subtable}
    % \end{adjustbox}
    % \caption{Values of the used constants}
    \label{tab:CharTimes}
\end{table*}
%%%%MAYBE a 2 column table%%%%%
%%%% MAKE TABLE PRETTY%%%%%

\subsubsection{Role of buoyancy}
Previous studies have shown that buoyancy may influence electroconvection \cite{DeValenca2017,Karatay2016}. Here we briefly discuss the role of buoyancy forcing in the present system. The relevant dimensionless parameter in this context is the Rayleigh number $Ra$, which compares buoyancy and viscous forces. For our configuration, we define
\begin{equation}
    Ra=\frac{g\beta d^4}{D_S\eta}\pdv{c}{z},
\end{equation}
where $g$ is the acceleration due to gravity, $d\approx \sqrt{D_St}$ is the characteristic length scale of the gradient and $\eta = 10^{-3}~Pa~s$ is the dynamic viscosity of water at 293 $K$. The density gradient can be estimated by using the concentration flux at the boundary $\pdv{c}{z} = i(1-\tau_{\text{H}^+})/(FD_S)$ in the electroneutrality limit \cite{Morris1963}, and the coefficient $\beta\approx \frac{\Delta \rho}{\Delta c}=0.0575$ kg/mol \cite{Brickwedde1949} for density change with depletion of $\text{HClO}_4$ at the electrode. With a critical value for the onset for buoyancy driven convection of $Ra_c = 817$ \cite{Sparrow1964,Tan1996} we obtain the relation 
\begin{equation}\label{eq:Ra1}
     it_B^2 = Ra_cF\eta/(g\beta(1-\tau_{\text{H}^+}))
\end{equation}
for the variation of the onset time $t_B$ for buoyancy convection. The implied dependence $t_B\sim 1/\sqrt{i}$ is compared to the other transition times in Fig. \ref{fig:constCurrent_2}(b). 

\citeauthor{Tan1996} \cite{Tan1996} have shown that the choice of the diffusion length scale (here $\sqrt{D_St}$) may under-predict the Rayleigh number and hence over-estimate $t_B$. We can check whether the estimate in equation \eqref{eq:Ra1} is accurate by re-defining the Rayleigh number using the numerical concentration profiles, similar to what was done by \citeauthor{DeValenca2017} \cite{DeValenca2017}. In Fig. \ref{fig:buoyancy}(a), we present the time evolution of the concentration profiles $c$ normalized with the initial/bulk concentration $c_0$ for the smallest applied $\abs{i}$. The effective diffusion layer thickness $\delta_B(t)$ is taken as the distance from the boundary where $c = 0.98c_0$, and we define the Rayleigh number now based on this quasi-steady length scale as
\begin{equation}
    Ra^*=\frac{g\beta \Delta c \delta_B^3}{D_S\eta},
\end{equation}
where $\Delta c = c_0-c(0,t)$ is the effective concentration depletion in the diffusion layer. The dotted line in Fig. \ref{fig:buoyancy}(a) indicates the estimated diffusion layer for numerical concentration profiles at three representative times.

The time evolution of $Ra^*$ at different $\abs{i}$ is  presented in Fig. \ref{fig:buoyancy}(b). Note that we have normalized the time-axis with Sand's time $t_{acid}$ (which is an estimate of electroconvection onset time from the model) to indicate whether buoyancy driven convection occurs before electroconvection. When $Ra^* > Ra_c$, the onset of buoyancy driven convection is expected. From Fig. \ref{fig:buoyancy}(b), we conclude that for the lower two current densities, the onset of buoyancy-mixing may be expected before electroconvection. This would explain the sudden smoothing of the pattern after its onset for the smallest current density (Fig. \ref{fig:ImgAllCur}) which is most likely caused by large-scale buoyant mixing. For the rest however, we observe that electroconvection precedes any density-driven mixing ($Ra(t_{acid})\ll Ra_c$). It therefore appears unlikely that the different scaling of the transition times for the three largest current densities, at $i/i_R\gtrapprox1$ in Fig. \ref{fig:constCurrent_2}(b), is a buoyancy effect. Nevertheless, since electroconvection can accelerate the onset of buoyancy driven convection \cite{Karatay2016}, it may drive mixing at later times. 

\subsubsection{Structure size and evolution} \label{sec:VortSize}

\begin{figure}
\includegraphics{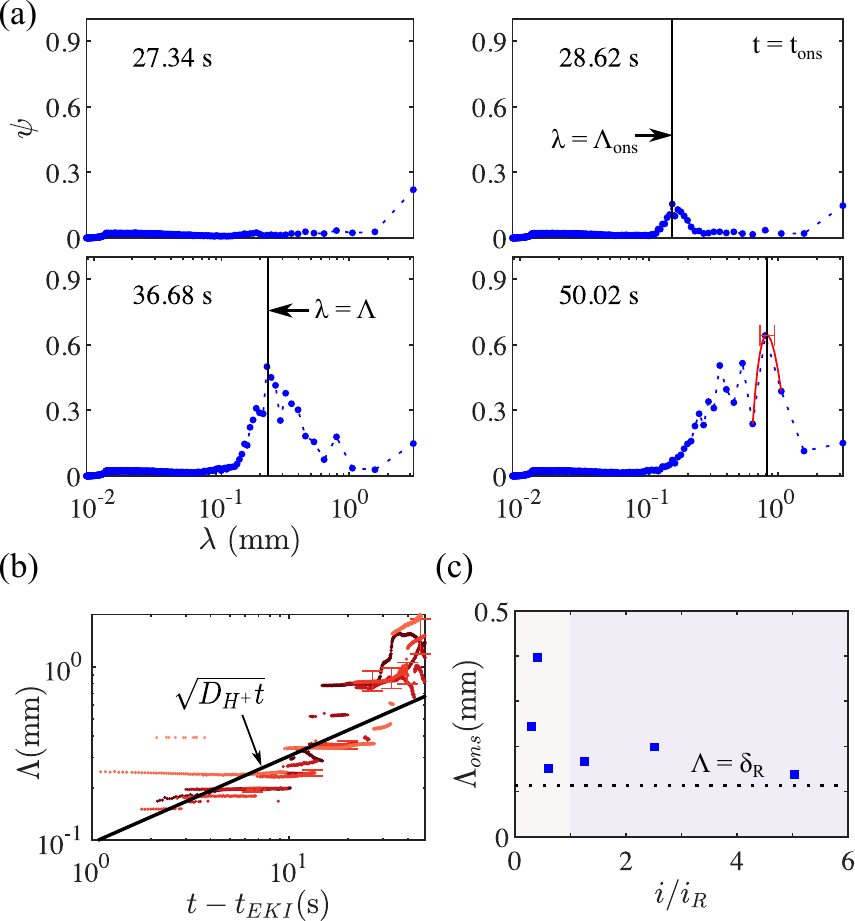}
\caption{\label{fig:waveL} (a)  Radially averaged power spectrum $\psi$, plotted against the wavelength $\lambda$ at different times for $\abs{i} = 0.95~\text{mA/cm}^2$. The dotted line is provided to guide the eye. A parabola (red curve) is fit around the maximum value and two adjacent points to obtain the wavelength of instability $\Lambda$. (b) The peak wavelength of the pattern $\Lambda$ is plotted as function of time, for all current densities after their onset. The color scheme is the same as used in Fig. \ref{fig:constCurrent_1}. Typical error bars have been shown for a single case $\abs{i} = 0.95~\text{mA/cm}^2$. The solid black line shows the diffusion length scale $\sqrt{D_{\text{H}^+}t}$. (c)  Structure size at onset $\Lambda(t_{ons}) = \Lambda_{ons}$ as a function of the current density, where $\delta_R$ is the size of the bulk reaction-diffusion boundary layer.}
\end{figure}

In addition to the transition times, we can also determine a typical length scale of the patterns as a measure of the size of the electroconvective vortices \cite{Yossifon2008}. We find the dominant wavelength $\Lambda$ of the patterns from the maximum in the radially averaged one-sided power spectrum $\psi$ of the mean subtracted fluorescence intensity of the images. In Fig. \ref{fig:waveL}(a) we present the time evolution of $\psi$ for a particular current density $\abs{i} = 0.95~\text{mA/cm}^2$. There is a clear dominant wavelength of the pattern which increases with time. At each time, $\Lambda$ is taken as the vertex of the parabola (red line) that is fit to the three points adjacent to the peak.

Fig. \ref{fig:waveL}(b) shows that when shifted by their respective instability time $t_{EKI}$, the growth of $\Lambda$ consistently follows a diffusive behavior with $\Lambda \approx \sqrt{D_{\text{H}^+}t}$ across all $\abs{i}$. The similarity in the pattern evolution at different current densities is also evident from fluorescence images directly (see Appendix \ref{sec:ImgCur}). This result is in line with the numerical prediction of \citeauthor{Rubinstein2000} \cite{Rubinstein2000} who showed that the size of electroconvective vortices in steady-state (which evolves through the merging of smaller vortices) is equal to the length scale of the diffusion domain. The agreement with the steady-state prediction, moreover, indicates that in our system, the vortices grow in a quasi-steady manner despite the transient evolution of the diffusion layer. Fig. \ref{fig:waveL}(b) is also consistent with measurements in IEMs \cite{Yossifon2008} where a similar scaling was measured for the electroconvective vortex size (for a single membrane pore, or nano-slot), although with a time-varying electrical forcing.

Note that with  $\sqrt{D_{\text{H}^+}t}$ being an estimate of the vertical propagation of the diffusive front, \sout{or} \rev{our} findings imply a typical aspect ratio (lateral wavelength over height of the structures) of 1. Lastly, in Fig. \ref{fig:waveL}(b) we plot the initial wavelength of the patterns $\Lambda_{ons} =  \Lambda(t_{ons})$ (taken at the empirical threshold value $\psi = 0.15$, see Fig. \ref{fig:waveL}(a) at $t \approx 29$ s). The value of $\Lambda_{ons}$ saturates at the reaction-diffusion boundary layer thickness $\Lambda = \delta_R$ for $i/i_R>1$, which is another manifestation of the limiting behavior observed for the pattern onset time.

\subsubsection{Convective motion}\label{sec:Conv}
\begin{figure}[b]
\includegraphics{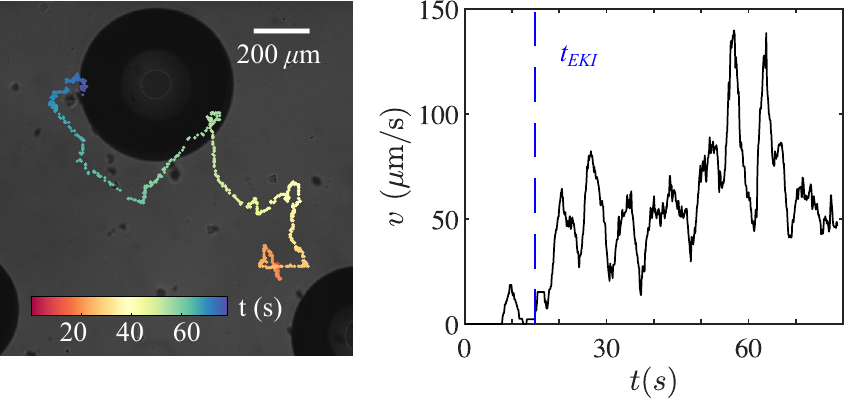}
\caption{\label{fig:convection} Measurements for $i \approx 2.3~\text{mA/cm}^2 =1.5~i_R$. Left: Time trace of the motion of a particulate impurity in solution. The particle path has been overlaid on the final (backlit) image taken of the electrode. The dark circle represents the bubble shadow. Right: In-plane velocity of the particle. $t_{EKI}$ has been measured from the fluorescence images. Full movie in supplementary material. Note that to obtain the time-trace and the velocity, the particle location was determined from each image frame by visual inspection.}
\end{figure}

In addition to the dye images and the footprint in the potential curves at $t_{EKI}$, we also obtain direct evidence of convective motion in the fluid by tracking the motion of a particulate impurity (size $\hat{a}\approx 40~\mu\text{m}$). The time trace of motion and the in-plane velocity are shown in Fig. \ref{fig:convection}. We find that the particle, initially at rest, moves vigorously only after $t = t_{EKI}$, with velocities, $v\sim 100~\mu \text{m/s}$, comparable to electroconvective velocities measured in \cite{DeValenca2017}. It should be noted that measurement with fluorescent tracer particles (Thermo Scientific, Fluoro-Max Red 36-2B, 6$\mu$m) failed, since presumably due to electrophoretic effects, these quickly disappeared from the measurement plane once the patterns emerged. The particulate impurity is apparently charged to a lesser extent compared to the tracers.  Yet it is still conceivable that also in this case the velocity $v_{EP} = \mu_p E_b$ induced by the bulk electric field $E_b$ may be significant. To test this, we estimate the electrophoretic mobility $\mu_p$ using the relation for a spherical colloidal particle \cite{Swan2012}: $\mu = \frac{2\varepsilon_r\varepsilon_0\zeta_p}{3\eta f(a/\lambda_D)}$, where $\varepsilon_0$ is the permittivity of vacuum, $\varepsilon_r = 80$ is the relative permittivity of water, $f(\hat{a}/\lambda_D)$ is Henry's function which depends on the ratio of the particle size to the Debye layer thickness $\lambda_D=\sqrt{\varepsilon_0 \varepsilon_r RT/(2F^2c_0)} = 6.8~\text{nm}$ (for $\hat{a}/\lambda_D\gg1$, $f = 1.5$) and $\zeta_p \approx 0.1~\text{V}$ is the assumed zeta-potential of the particle. We approximate $E_b$, based on the steady-state diffusion of counter ions as $E_b = \frac{RT/F}{c}\pdv{c}{z}\approx\frac{RT/F}{\delta_R} = 220~\text{V/m}$, where diffusion length is taken to be its limiting value $\delta_R$. The electrophoretic velocity is therefore calculated to be $v_{EP}\approx16~\mu \text{m/s}$. $v_{EP}$, although smaller than the maximum particle velocity $v_{max}\approx 150~\mu\text{m/s}$, is of the same order of magnitude to the value measured soon after pattern onset ($\approx 50~\mu\text{m/s}$). This implies that the electrophoretic force on the particle cannot be disregarded and that an accurate measurement of the fluid velocity would require particles with well defined zeta potentials. Nevertheless, since $v_{max}$ is an order of magnitude larger $v_{EP}$, it is likely that there is fluid motion in the present system. 

A final remark is regarding the roles of the bubbles that can be seen to nucleate occasionally (see Fig. \ref{fig:CV}) and can cause bending of electric-field lines around them (the case of a dielectric sphere in an electric-field \cite{Schnitzer2014}). However, the presence of such a perturbation, does not appear to significantly influence the transition times as evidenced by the good agreement between the experimental results and those from the 1-D ion transport model. Further, the pattern formation does not necessarily initiate from bubble locations and the structures generally appeared unaffected by the bubbles, except in their immediate vicinity.

\section{Conclusion}\label{sec:conclusion}

In summary, we have investigated electroconvective pattern formation in water electrolysis and have linked these findings to previous measurements \rev{of electroconvective phenomena} on ion-exchange membranes. \rev{The measured characteristic times differ between the two systems only at large current densities ($i>i_R$), due to the bulk chemical reaction for water equilibrium. Our estimates suggest that buoyancy plays a role in the present system, especially at the lower current densities. However, density driven convection is not likely to influence the deviating trends of the transition times observed for larger $i$. We further present current and voltage measurements, and have compared them to estimates from simple models. These results along with pattern images may prove useful as a benchmark for future modelling and simulation efforts of electroconvection in electrolytic systems.} \rev{Finally, we also} \sout{We additionally} reveal pattern formation up to a supporting electrolyte concentration  $c_{sup}=1$ mM, i.e for $c_{sup}/c_{\text{H}^+} \leq 0.5$ (see Appendix \ref{sec:ImgSalt}).

While electroconvection is unlikely in commercial electrolyzers (where excess supporting electrolyte is used), these findings are relevant to laboratory electrochemical studies which are frequently performed in the absence of supporting electrolyte \cite{Sheng2010,Bashkatov2019}. Moreover, electroconvection could be the possible driving mechanism for the ``spontaneous convection'' (different from buoyancy driven convection) which is assumed by \citeauthor{Amatore2001} to successfully fit experimentally measured currents \cite{Amatore2001}. Measurements similar to those presented in this paper, with charged fluorescent dyes, could help check for this possibility. \rev{Note, that the relation of overlimiting currents to electroconvective phenomena shown in the present study is only valid for planar electrode geometries. For more practical electrode designs, such as porous electrodes, or membrane-electrode geometries (which may allow gas exchange and have a large surface area), overlimiting currents could also be driven by the interaction of the surface charge of pore walls with the electric field (surface conduction and electro-osmotic flow \cite{Dydek2011}). This may especially be relevant when the catalyst layer is separate from the pores, where there might also be a concentration gradient along the pore length.}

\sout{However, what} \rev{It} is readily apparent \rev{from the present study}, \sout{is,} that in order to explain the current or voltage response of electrolytic cells, a complete understanding of the relevant ion transport processes is needed. This includes recognizing the role of convective instabilities and non-linear bulk reactions in \sout{electrochemical} \rev{electrolytic} systems. \rev{Additional experimental measurements under well controlled conditions: with simple geometry (e.g. imposed 1D flow) and reduced number of interacting ions (e.g. in $\text{CO}_2$ reduction with and without a buffer electrolyte), would certainly help establish to what extent transport processes affect the potential drop in solution.}

% The work presented here, existence of electroconvective phenomenon in water electrolysis.

% \rev{We hope that our current or voltage measurements, obtained for the analytically simple case of a binary acidic electrolyte, can be useful in validating the choice of particular models of electrolysis. The pattern images could also be used as a benchmark for future modelling an simulation efforts of electroconvection in electrolytic systems. Additional experimental measurements in electrolytic systems under well controlled conditions, for example with simple geometry and reduced number of interacting ions (ex: in $\text{CO}_2$ reduction without a buffer electrolyte) would certainly be useful in achieving the end goal: predicting the current/voltage response of electrolytic cells.}

\appendix
\renewcommand\thefigure{\thesection.\arabic{figure}}  

\section{Comparison of measured and theoretical Linear Sweep Voltammograms}\label{sec:LSV_norm}
\setcounter{figure}{0}

\rev{To further understand the $i-\phi$ measurement, a comparison could be made with an analytic solution of current density for linear sweeps: for example a linear sweep with fast, reversible electrode kinetics (i.e. obeying the Nernst equation) \cite{BardFaulkner2002, Yan2017}. A plot of the analytical expressions usually reveals a transient peak in the current density at the potentials at which the electrode reaction reaches diffusion limitation \cite{BardFaulkner2002, Yan2017}. This is similar to the first green marker in Fig. \ref{fig:CV}(b).} 
%and is obtained when the potential sweep time scale is much faster than the rate of diffusion, i.e. $\frac{vL^2}{D\phi_T}\gg 1$. Here $L$ is the system length scale, $D$ is the diffusion coefficient of the reacting ion and $\phi_T = RT/F \approx 25 mV$ is the thermal voltage ($R$ is the ideal gas constant, $T$ = 298 K). For a semi-infinite domain ($L \rightarrow \infty$) this is always the case.
\rev{Following this peak, the theoretical curve decreases and approximately follows the Cottrell equation, i.e. approaching $i_{lim}$ \footnote{This result can be easily shown by plotting, for example, the modified Randles-Sevcik equation \cite{Yan2017} (equation 33 there) alongside $i_{lim}$ in Fig. \ref{fig:CV}(b). However, since it does not add substantially to the figure, this is avoided here.}. We consider one such analytical expression, the modified Randles-Sevcik equation obtained by \citeauthor{Yan2017} \cite{Yan2017} (equation 33 there) which is valid for fast reactions (Nernst equation valid at the electrode boundary) in a fully supported electrolyte (no potential drop in the electrolyte). It therefore involves solving the semi-infinite diffusion equation for a single reacting ion. Note however that such an estimate cannot directly be compared with the $i-\phi$ measurement in Fig. \ref{fig:CV}(b) (black line). Since the analytical estimate neglects any potential drop in the bulk solution, in contrast to the experimental measurement (for the same potential sweep rate), it reaches the potential of diffusion limitation (the potential at which $c_{\text{H}^+} = 0$ at the electrode) much sooner. The thickness of the diffusion layer $\sim \sqrt{D_{\text{H}^+}t}$ is therefore much smaller, and the predicted current density much larger than the experiment.}

\begin{figure}
\includegraphics{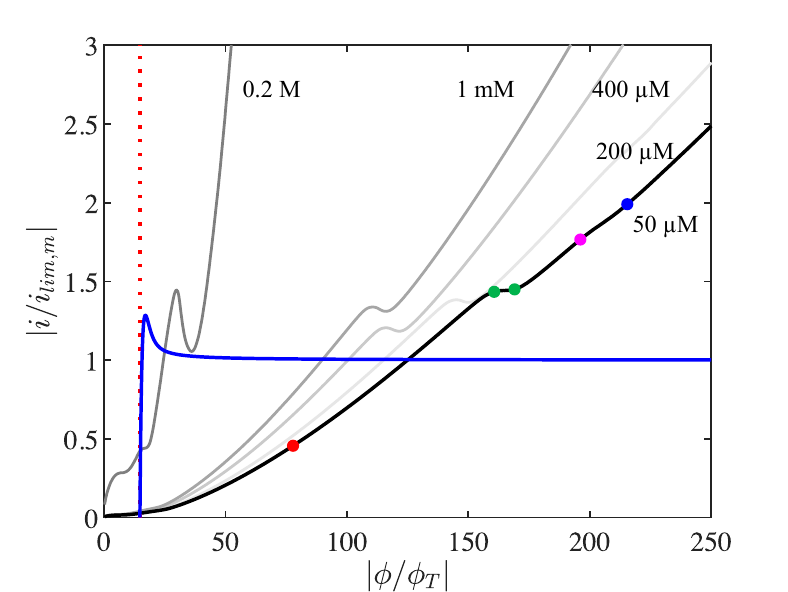}
\caption{\label{fig:CVNormalized}\rev{Representative plot showing the normalized Linear sweep voltammogram measured for different mentioned supporting electrolyte concentration for scan rate $v = 0.1~\text{V/s}$. The blue line is the analytical estimate and is based on the modified Randles-Sevcik equation (equation 33 in \cite{Yan2017}), applicable for fast-reactions in a fully supported solution (infinite supporting electrolyte) involving only one ionic species. $\phi = \phi_{rev}$ is taken as the x-ordinate of the estimate and is given here by the red dotted line.}}
\end{figure}

\rev{We scale out these differences in current densities by normalizing the analytical estimate and measurement with their respective Cottrell-like relationships. This result is presented in Fig. \ref{fig:CVNormalized} where the primary measurement (black line) is compared with the analytical estimate (blue line). We also include results of linear sweep measurements done in solutions with different quantities of supporting electrolyte. Specifically, each $i$ is normalized with a Cottrell-like current density $i_{lim,m}$:
% We plot this result in Fig. \ref{fig:CVNormalized}, and compare it with $i-\phi$ curve presented in the main text (Fig. \ref{fig:CV}(b), black line). We also include results of linear sweep measurements done in solutions with different quantities of supporting electrolyte. 
% To scale out any change in current brought out only due to the growing diffusion boundary, i.e. $i\sim 1/\sqrt{Dt}$, 
% we normalize each $i$ with a Cottrell-like current density $i_{lim,m}$:
\begin{equation}
    i_{lim,m}(t) = \rev{-FD_{11}\frac{c_0}{(1-\tau_{\text{H},m})\sqrt{\pi D_{11}t}}}, \label{cottrell_Supp}
\end{equation}
Equation \eqref{cottrell_Supp} is the modified limiting current density, $i_{lim,m}$, which is a semi empirical extension of equation \eqref{cottrell} for different supporting electrolyte concentrations \cite{Morris1963}.
The modified diffusion coefficient, $D_{11}$, and modified ion transport number $\tau_{\text{H,m}}$ are given by \cite{Morris1963},
\begin{subequations}
\begin{align}
    D_{11} &= D_{\text{H}^+}\left(1- \frac{c_0(D_{\text{H}^+}-D_{\text{ClO}_4^-})}{c_0(D_{\text{H}^+}+D_{\text{ClO}_4^-}) + c_{sup}(D_{\text{Na}^+}+D_{\text{ClO}_4^-})}\right)\\
    \tau_{\text{H,m}} &= \frac{c_0D_{\text{H}^+}}{(c_0D_{\text{H}^+} + (c_{sup}+c_0)D_{\text{ClO}_4^-} + c_{sup}D_{\text{Na}^+})}
\end{align}
\end{subequations}
Note that for $c_{sup} \gg c_0$, $D_{11} = D_{\text{H}^+}$ and $\tau_{\text{H,m}} = 0$ and equation \eqref{cottrell_Supp} reduces to the usual form of the Cottrell equation \cite{BardFaulkner2002}. For the other extreme case of a binary electrolyte, when $c_{sup} \ll c_0$, $D_{11} = D_{S}$ and $\tau_{\text{H,m}} = \tau_{\text{H}^+}$, we obtain equation \eqref{cottrell}. Thus, normalizing with the modified Cottrell equation allows better comparison between the $i-\phi$ relationships for all supporting electrolyte concentrations, since it helps scale out the differences in bulk ion transport.}

\rev{It is apparent from Fig. \ref{fig:CVNormalized} that for all supporting electrolyte concentrations, there is a transient peak associated with the fast linear sweep. There is also a clear trend: for the lower conductivity solutions, a larger potential is required to achieve diffusion limitation, likely because of greater electrical resistance of the electrolyte. The analytical result, valid for purely diffusive transport, further shows that $i$ decreases to the diffusion limited value $i_{lim,m}$ soon after the peak current. Comparing this curve with the experimental measurements, it can be seen that overlimiting currents are driven in all cases beyond a point. For a binary electrolyte (here $c_{sup} = 50~\mu M$) the overlimiting current is associated with pattern formation and electroconvection, as shown in the main text. With increasing $c_{sup}$, pattern formation (and likely electroconvection) is observed much later, and not at all for $c_{sup} = 0.2~M$. In these scenarios it is likely that buoyancy driven mixing (due to gradients in the concentration of the supporting electrolyte), supports the overlimiting current transport in the bulk electrolyte.}

\rev{It should be noted that in the main text equation \eqref{cottrell} is applied from the time when $\abs{\phi} > \abs{\phi_{rev}}$. This is not the case when calculating $i_{lim,m}$ for the measured curves here. The time $t = 0$ in $i_{lim,m}$ for the experimental measurements, corresponds to $\abs{\phi} = 0~\text{V}$ vs Ag/AgCl (and not $\phi = \phi_{rev} = 0.3972~\text{V}$) (this is done so that $i$ values at $\phi<\phi_{rev}$ are also normalized). In contrast, the analytical expression (and corresponding $i_{lim,m}$) only begins at $\phi = \phi_{rev}$. This however does not change the discussion and conclusion drawn in the preceding paragraphs.}

\section{Numerical model}\label{sec:numerics}
\setcounter{figure}{0}  

A simplified numerical model is derived here, primarily to calculate Sand's time for the present system.

The general form of the Nernst-Planck-Poisson equations with a bulk reaction is:
\begin{subequations}
\begin{align}
\pdv{c_k}{t} &= \pdv{J_k}{z}+\text{R}_k,\\
J_k &= D_k\left(\pdv{c_k}{z} + \frac{\zstroke_kFc_k}{RT}\pdv{\phi}{z}\right),\\
\varepsilon\laplacian \phi  &= -\sum_{k=1}^{n}\zstroke_kFc_k \label{eq:Poisson}
\end{align}
\label{eq:first}
\end{subequations}
\noindent where $c_k$ is the concentration, $J_k$ is the mass flux and $\zstroke_k$ is the sign of the charge of k'th ionic species, for a total `$n$' number of ions. $\text{R}_k$ is the bulk reaction for each equation, $\phi$ is the electrostatic potential in solution and $\varepsilon$ is the absolute permittivity of water. More specifically, taking an acid with anion $\text{ClO}_4^-$ and a supporting salt with cation $\text{Na}^+$ and anion $\text{ClO}_4^-$, we have:
\begin{subequations}
\begin{align}
    &\pdv{c_{\text{H}^+}}{t} = D_{\text{H}^+}\pdv{}{z}\left(\pdv{c_{\text{H}^+}}{z} + \frac{Fc_{\text{H}^+}}{RT}\pdv{\phi}{z}\right)+R\label{eq:H}\\
    &\pdv{c_{\text{ClO}_4^-}}{t} = D_{\text{ClO}_4^-}\pdv{}{z}\left(\pdv{c_{\text{ClO}_4^-}}{z} - \frac{Fc_{\text{ClO}_4^-}}{RT}\pdv{\phi}{z}\right)\label{eq:ClO}\\
    &\pdv{c_{\text{OH}^-}}{t} = D_{\text{OH}^-}\pdv{}{z}\left(\pdv{c_{\text{OH}^-}}{z} - \frac{Fc_{\text{OH}^-}}{RT}\pdv{\phi}{z}\right)+R\label{eq:OH}\\
    &\pdv{c_{\text{Na}^+}}{t} = D_{\text{Na}^+}\pdv{}{z}\left(\pdv{c_{\text{Na}^+}}{z} + \frac{Fc_{\text{Na}^+}}{RT}\pdv{\phi}{z}\right)\label{eq:Na}
\end{align}
    \label{eq:potGrad}
\end{subequations}
Here, $R = k_bc_W-k_fc_{\text{H}^+}c_{\text{OH}^-}$ is the bulk dissociation of water for the reaction,
\begin{align}
    \ce{  H+ + OH- &<=>[k_f][k_b] H2O}\\
\end{align}
where, $k_b = 2.6\times10^{-5}$ $s^{-1}$ is the water dissociation rate constant, and $k_f = 1.4\times10^{11}$ $\text{M}^{-1}s^{-1}$ the rate constant for $\text{H}^+$ and $\text{OH}^-$ association \cite{Paldus1976b}. Note that the concentration of water, $c_W = 55.55$ M, is assumed to be constant (and much larger than the $\text{H}^+$ or $\text{OH}^-$ concentration.

\subsection{(Acidic) binary electrolyte (acid)}\label{sec:acid}
Assuming electroneutrality, we get continuity of the current i.e the current (or charge flux) is constant in the system and equal to the value at the boundary. In other words,
% \begin{equation}
% \begin{align}
% \begin{split}
\begin{multline}
\Sigma~\zstroke_k J_k = \pdv{}{z}\left(D_{\text{H}^+}c_{\text{H}^+} - D_{\text{ClO}_4^-}c_{\text{ClO}_4^-} \right. \\ 
\left. - D_{\text{OH}^-}c_{\text{OH}^-} + D_{\text{Na}^+}c_{\text{Na}^+}\right)\\
+ \frac{F}{RT}\pdv{\phi}{z}\left(D_{\text{H}^+}c_{\text{H}^+} + D_{\text{ClO}_4^-}c_{\text{ClO}_4^-} \right. \\ + D_{\text{OH}^-}c_{\text{OH}^-} + D_{\text{Na}^+}c_{\text{Na}^+})= \frac{i}{F}
\end{multline}
% \end{split}
% \end{align}
% \end{equation}
Since electroneutrality implies that $c_{\text{H}^+} + c_{\text{Na}^+} = c_{\text{ClO}_4^-} + c_{\text{OH}^-} = c$, the above equations can be rewritten as

\begin{multline}
\pdv{}{z} \left( (D_{\text{H}^+}-D_{\text{OH}^-})c - (D_{\text{H}^+}-D_{\text{Na}^+})c_{\text{Na}^+}\right. \\ \left. + (D_{\text{OH}^-} - D_{\text{ClO}_4^-})c_{\text{ClO}_4^-} \right) \\
+ \frac{F}{RT}\pdv{\phi}{z} \left( (D_{\text{H}^+}+D_{\text{OH}^-})c - (D_{\text{H}^+}- D_{\text{Na}^+})c_{\text{Na}^+} \right. \\ \left. -(D_{\text{OH}^-}-D_{\text{ClO}_4^-})c_{\text{ClO}_4^-} \right) = \frac{i}{F}
\end{multline}

Consider the case where $c_{\text{Na}^+},c_{\text{OH}^-}\ll c_{\text{H}^+}=c_{\text{ClO}_4^-} = c$, i.e. the (initial) concentration of $\text{OH}^-$ is very small and there is close to no supporting electrolyte in solution. The gradient in the potential in this case is given by

\begin{equation}
\pdv{\phi}{z}= \frac{RT}{Fc}\frac{1}{(D_{\text{H}^+} + D_{\text{ClO}_4^-})}\left(\frac{i}{F}-(D_{\text{H}^+}-D_{\text{ClO}_4^-})\pdv{c}{z}\right)\label{eq:phiGrad}
\end{equation}
Furthermore, note that in equation \eqref{eq:potGrad} the contribution of the potential gradient term is of the form
\begin{equation}
\begin{split}
    &D_k\pdv{}{z}\left(c_k\pdv{\phi}{z}\right) = \\&D_k\pdv{}{z}\left(\frac{c_k}{c}\frac{RT}{F(D_{\text{H}^+}+D_{\text{ClO}_4^-})}(i-(D_{\text{H}^+}-D_{\text{ClO}_4^-}) \pdv{c}{z})\right)
\end{split}
\end{equation}

Since $c_{\text{OH}^-}/c \sim c_{\text{Na}^+}/c \sim 0$, the contribution of the potential gradient term to the transport of $\text{OH}^-$ and the salt cation is negligible. Equation \eqref{eq:OH} and \eqref{eq:Na} can now be written as:
\begin{subequations}
    \begin{align}
    &\pdv{c_{\text{OH}^-}}{t} = D_{\text{OH}^-}\pdv{}{z}\left(\pdv{c_{\text{OH}^-}}{z}\right)+\text{R},\\
    &\pdv{c_{\text{Na}^+}}{t} = D_{\text{Na}^+}\pdv{}{z}\left(\pdv{c_{\text{Na}^+}}{z}\right)
    \end{align}\label{eq:Na1}
\end{subequations}
with the boundary conditions:
\begin{subequations}
\begin{align}
    c_{k}\eval_{ z\rightarrow \infty} = c_{k,initial}
\end{align}
\end{subequations}
and
\begin{subequations}
\begin{align}
    D_{\text{OH}^-}\left(\pdv{c_{\text{OH}^-}}{z}\right)\eval_{z=0} &= 0,\\
    D_{\text{Na}^+}\left(\pdv{c_{\text{Na}^+}}{z}\right)\eval_{z=0} &= 0. \label{eq:NaBound}
\end{align}
\end{subequations}

Also, equation \eqref{eq:H} and \eqref{eq:ClO} and the related boundary conditions,
\begin{subequations}
\begin{align}
     D_{\text{H}^+}\left(\pdv{c}{z} + \frac{Fc}{RT}\pdv{\phi}{z}\right)\eval_{z=0} &= \frac{i}{F},\label{eq:HBoundary}\\
    D_{\text{ClO}_4^-}\left(\pdv{c}{z} - \frac{Fc}{RT}\pdv{\phi}{z}\right)\eval_{z=0} &= 0
    \end{align}
\end{subequations}
can further be reduced to the diffusion equation, with the additional reaction term, for an effective salt species $c_S$, following the procedure for a binary electrolyte by \cite{Morris1963, Newman2004}.

The final set of equations that are solved are:
\begin{subequations}
    \begin{align}
    \pdv{c_S}{t} &= D_S\pdv[2]{c}{z} + \frac{D_{\text{ClO}_4^-}}{D_{\text{H}^+}+D_{\text{ClO}_4^-}}\text{R}\\
    \pdv{c_{\text{OH}^-}}{t} &= D_{\text{OH}^-}\pdv[2]{c_{\text{OH}^-}}{z}+\text{R}
    \end{align}
\end{subequations}
with boundary conditions
\begin{subequations}
\begin{align}
    c_{S}\eval_{z = l} & = c_0,\\
    c_{\text{OH}^-}\eval_{z = l} & = \frac{k_bc_W}{k_fc_0},
\end{align}
\end{subequations}
and, 
\begin{subequations}
\begin{align}
     \frac{D_S}{1-t_H}\pdv{c_S}{z}\eval_{z = 0} &= \frac{i}{F},\label{eq:BC1}\\
    D_{\text{OH}^-}\left(\pdv{c_{\text{OH}^-}}{z}\right)\eval_{z = 0} &= 0,
\end{align}
\end{subequations}
where, $D_S = 2D_{\text{H}^+}D_{\text{ClO}_4^-}/(D_{\text{H}^+}+D_{\text{ClO}_4^-})$ and $\tau_{\text{H}^+} = D_{\text{H}^+}/(D_{\text{H}^+}+D_{\text{ClO}_4^-})$.

\begin{figure}
\includegraphics{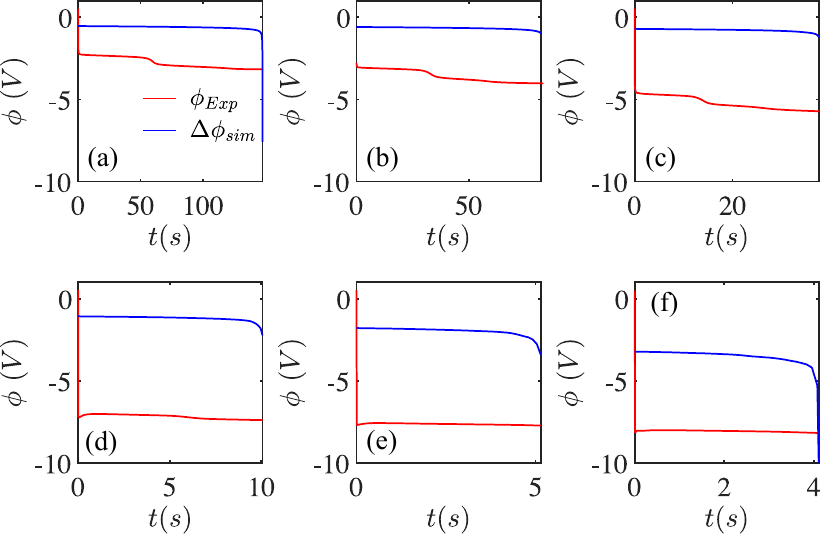}
\caption{\label{fig:phi} The measured electrode potential $\phi_{Exp}$ compared to that calculated from the electroneutral numerical model $\Delta \phi_{sim}$. The difference indicates the potential drop in the Debye layer.}
\end{figure}

\subsection{Excess supporting electrolyte (supp)}
We can also derive a simplified model for ion transport for the case when the electrolyte solution contains excess supporting electrolyte. While not directly applicable to the present experimental system, it helps provide a lower limit for the measured transition times.
Here, besides assuming electroneutrality, we also have $c_{\text{H}^+},c_{\text{OH}^-} \ll c_{\text{Na}^+}=c_{\text{ClO}_4^-} = c$. Similarly, the potential gradient is given by

\begin{equation}
\pdv{\phi}{z}= \frac{RT}{Fc}\frac{1}{(D_{\text{Na}^+} + D_{\text{ClO}_4^-})}\left(\frac{i}{F}-(D_{\text{Na}^+}-D_{\text{ClO}_4^-})\pdv{c}{z}\right)
\end{equation}
Here, instead $c_{\text{OH}^-}/c \sim c_{\text{H}^+}/c \sim 0$, and the contribution of the potential gradient term to the transport of $\text{H}^+$ and $\text{OH}^-$ is negligible. The equations reduce to a reaction-diffusion system
\begin{subequations}
\begin{align}
    \frac{\partial{c_{\text{H}^+}}}{\partial{t}} &= D_{\text{H}^+}\frac{\partial{c_{\text{H}^+}}}{\partial{z^2}} + \text{R},\\
    \frac{\partial{c_{\text{OH}^-}}}{\partial{t}} &= D_{\text{OH}^-}\frac{\partial{c_{\text{OH}^-}}}{\partial{z^2}} + \text{R}
\end{align}
\end{subequations}
with boundary conditions,
\begin{subequations}
\begin{align}
    c_{\text{H}^+}(l) & = c_0,\\
    c_{\text{OH}^-}(l) & = \frac{k_bc_W}{k_fc_0}
\end{align}
\end{subequations}
at $z = l$ (where $c_0$ is the initial $\text{H}^+$ concentration) and
\begin{subequations}
\begin{align}
    \frac{\partial{c_{\text{H}^+}}}{\partial{z}} & = \frac{i_F}{D_{\text{H}^+}F},\\
    \frac{\partial{c_{\text{OH}^-}}}{\partial{z}} & = 0
\end{align}
\end{subequations}
at the electrode surface. Note that in this case, the potential gradient in solution is obtained by solving the transport equations for the supporting electrolyte.

\begin{table}
 \begin{adjustbox}{max width=\columnwidth}
    \begin{tabular}{|l|l|}
    \hline
     $D_{\text{H}^+} = 9.3\times10^{-9}$ $\text{m}^2/\text{s}$ \cite{Robinson1959} & $D_{\text{OH}^-} = 4.5\times10^{-9}$ $\text{m}^2/\text{s}$ \cite{Daniele2002}\\
     \hline
     $k_b = 2.6\times10^{-5}$ $\text{s}^{-1}$ \cite{Paldus1976b}  & $k_f = 1.4\times10^{11}$ $\text{s}^{-1}\text{M}^{-1}$ \cite{Paldus1976b}\\
     \hline
     $D_{\text{ClO}_4^-}= 1.792\times10^{-9}$ $\text{m}^2/\text{s}$ \cite{Heil1995} & $c_W = 55.55~\text{M}$\\
     \hline
    \end{tabular}
    \end{adjustbox}
    \caption{Values of the used constants}
    \label{tab:ConstVal}
\end{table}

We use a second order finite difference approximation with the integrating factor (IIF) numerical scheme presented in \citeauthor{Nie2006} \cite{Nie2006} to solve the reaction-diffusion models above (as done in in our recent work \cite{Pande2020}). For the calculation, we use a time resolution of $\Delta t = 0.03$ s and a spatial resolution of $\Delta z = 10 \mu\text{m}$ for the calculations. Values of the constants used in presented in Table \ref{tab:ConstVal}. The corresponding Sand's time for the above two models, $t_{acid}$ and $t_{supp}$, is the time when the concentration of the reacting ion $c=0$ at the boundary.

\subsection{Estimating potential drop in the double layer}\label{sec:PotDeb}

For the chronopotentiometric experiments, the measured potential $\phi_{Exp}$ is the total potential drop between the working and the reference electrode, i.e.  $\phi_{Exp} = \Delta \phi_S + \Delta \phi_D + \Delta \phi_{bulk}$, where $\Delta \phi_S$ and $\Delta \phi_D$ are the potential drop in Stern and diffuse layer respectively, and $\Delta \phi_{bulk}$ is the potential drop in the bulk electroneutral solution. $\Delta \phi_S$ is the potential difference that drives the reaction, and is assumed to be equal to the reversible potential for hydrogen evolution ($0~V$ vs RHE). Measured on the Ag/AgCl reference electrode scale, this value is $\phi_{rev} = 0 - 0.059\text{pH} - \phi^0_{Ag/AgCl} = \rev{-}0.3792~\text{V}$ where $\phi^0_{Ag/AgCl} = 0.22~\text{V}$ (at $298~K$) is the standard potential for the reference electrode \cite{Bates1978}. Furthermore, the results of the numerical model '$acid$' can be used to calculate $\Delta \phi_{bulk}$ by integrating equation \eqref{eq:phiGrad}. This combined numerical value $\Delta \phi_{sim} = \Delta \phi_{bulk} + \phi_{rev}$ is presented along with the $\phi_{Exp}$ in Fig. \ref{fig:phi} (a)-(e) for increasing $\abs{i}$ respectively. The difference is an estimate of $\Delta \phi_D$, where for all current densities $\Delta \phi_D>1~\text{V}$. This implies that within the Debye layer which has a thickness of $\lambda_D = 6.8~\text{nm}$, the electric field strength $E_D \gtrapprox 1/\lambda_D ~\text{V/m}= 1~\text{MV/cm}$. This is sufficiently large to increase the dissociation rate of water \cite{Onsager1934,Tanaka2010} beyond its bulk value $k_b$. For $E_D=1~\text{MV/cm}$ the increased dissociation constant is calculated to be $\approx 3k_b$ (using equation (38) in \cite{Tanaka2010}). Thus the water dissociation rate is at least three times as fast in the double layer than in the bulk solution and can plausibly contribute additional protons for the reaction at the electrode.

\section{Patterns at different supporting electrolyte concentrations}\label{sec:ImgSalt}
\setcounter{figure}{0}  

\begin{figure}
\includegraphics{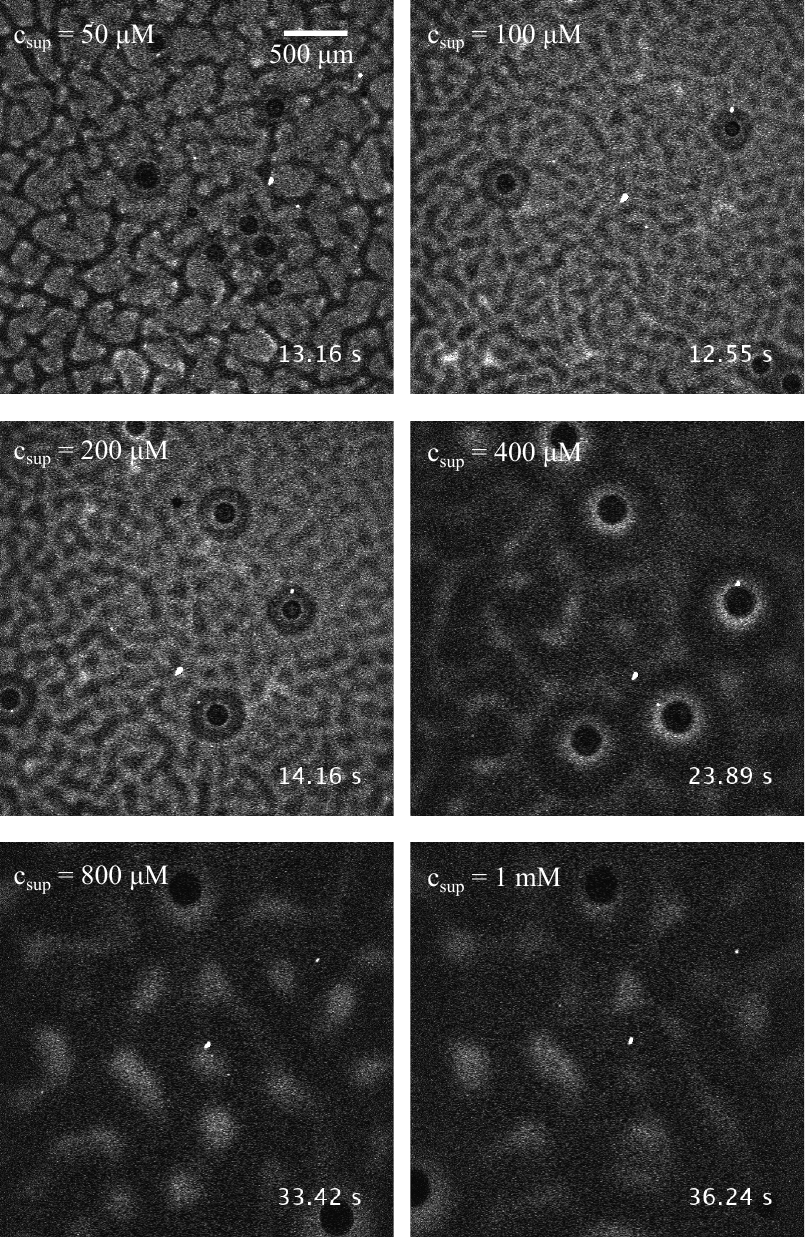}
\caption{\label{fig:patternSup} Panels showing the patterns formed for different supporting electrolyte concentration (shown in the upper left corner) for $\abs{i} = 7.96~\text{mA/cm}^2$ at their onset. The time of onset of the pattern, after the current is applied, is shown in the bottom right corner.}
\end{figure}

In Fig. \ref{fig:patternSup} we show additional images of patterns at onset (visual inspection), at different supporting salt concentrations (here $\text{NaClO}_4$; $c_{\text{sup}}$), for the highest current density considered in our work $i = 7.96~ \text{mA/cm}^2$ ($i \approx 5 i_R $). In each of the case, $c_{\text{Na}^+}<c_{\text{H}^+}$. The images show the pattern at the moment of onset, and the time of onset is mentioned in the right bottom corner of the image. A general observation is that with increasing supporting electrolyte concentration, $c_{sup}$, the onset time of the pattern as well as the initial wavelength of the pattern (wavelength of fluorescence intensity variation) increases. There appears to be, in fact, a sudden increase in initial pattern wavelength between $ 200~\mu\text{M}<c_{sup}< 400~\mu\text{M}$. 

\section{Panels of pattern images for all current densities}\label{sec:ImgCur}
\setcounter{figure}{0}  

\begin{figure*}
\includegraphics[scale = 0.92]{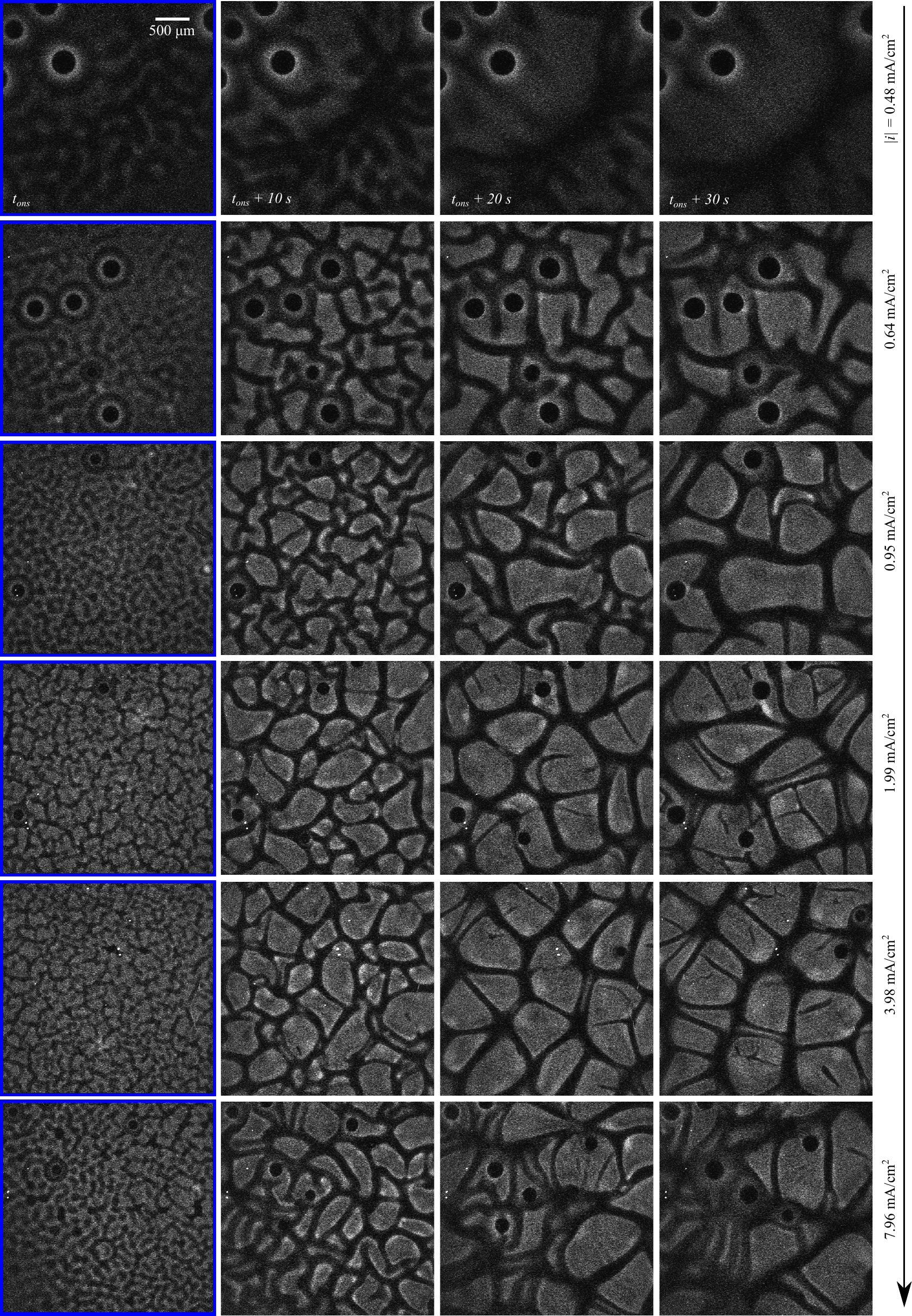}
\caption{\label{fig:patternI} Panels showing the patterns formed for different applied currents at $\approx 10 s$ intervals after onset $t_{ons}$.}
\label{fig:ImgAllCur}
\end{figure*}

In Fig. \ref{fig:patternI} we present the images of the pattern at equal intervals (10 s) after their onset $t_{EKI}\approx t_{ons}$. The pattern images for almost all the current densities appears to grow in a similar manner as shown quantitatively in Fig. 2(d) in the main text. For the lowest current density however the smoothing of the pattern could be brought about by large scale buoyant convection. 
\clearpage
\begin{acknowledgments}
This research received funding from The Netherlands Organization for Scientific Research (NWO) in the framework of the fund New Chemical Innovations, project ELECTROGAS (731.015.204),
with financial support of Akzo Nobel Chemicals, Shell Global Solutions, Magneto Special Anodes (an Evoqua Brand), and Elson Technologies. We acknowledge The Netherlands Center for Multiscale Catalytic Energy Conversion (MCEC) and the Max Planck Center Twente for Complex Fluid Dynamics for financial support. D.L. also acknowledges financial support by an ERC-Advanced Grant.
\end{acknowledgments}

\bibliography{./bib/Patterns.bib}

\end{document}